\RequirePackage{fix-cm}

\documentclass[smallextended]{svjour3}       

\smartqed  

\usepackage{url}

\usepackage[hidelinks]{hyperref}
\AtBeginDocument{}%

\usepackage{multirow}

\usepackage{graphics,epsfig,epsf,rotating}

\usepackage{rotfloat,array,pifont}

\usepackage[numbers]{natbib}

\usepackage{pdflscape}

\usepackage{color}
\ifx\release\undefined
    \def\printcoments{1}
\fi
\ifx\printcoments\undefined
            \newcommand{\comm}[1]{}
            \newcommand{\commatd}[1]{}
\else
            \newcommand{\comm}[1]{\textbf{{\textcolor{red}{#1}}}}
            \newcommand{\commatd}[1]{\textbf{{\textcolor{blue}{#1}}}}
\fi

\begin{document}

\title{Understanding Federation: An Analytical Framework for the Interoperability of Social Networking Sites
}

\titlerunning{An Analytical Framework for the Interoperability of SNS}        

\author{Antonio Tapiador         \and
        Samer Hassan 
}


\institute{Antonio Tapiador \at
              Universidad Polit\'{e}cnica de Madrid, c/ Ramiro de Maeztu, 7, 28040 Madrid, Spain \\
              \email{atapiador@dit.upm.es} \\
              Universidad Complutense de Madrid, c/ Profesor Jos\'{e} Garc\'{i}a Santesmases, 9, 28040 Madrid, Spain \\
              \email{atapiador@ucm.es}           
           \and
           Samer Hassan \at
              GRASIA research group, Universidad Complutense de Madrid, c/ Profesor Jos\'{e} Garc\'{i}a Santesmases, 9, 28040 Madrid, Spain \\
              \email{samer@fdi.ucm.es} 
}

\date{}

\maketitle

\begin{abstract} 

Although social networking has become a remarkable feature in the Web, full interoperability has not arrived. This work explores the main 5 paradigms of interoperability across social networking sites, corresponding to the layers in which we can find interoperability. Building on those, a novel analytical framework for SNS interoperability is introduced. Seven representative interoperability SNS technologies are compared using the proposed framework. 
The analysis exposes an overwhelming disparity and fragmentation in the solutions for tackling the same problems. Although there are a few solutions where consensus is reached and are widely adopted (e.g. in object IDs), there are multiple central issues that are still far from being widely standarized (e.g. in profile representation). In addition, several areas have been identified where there is clear room for improvement, such as privacy controls or data synchronization. 

\end{abstract}

\section{Introduction}



Although social networking has become a remarkable feature in the Web, full interoperability has never arrived.
Web users develop their personal and professional activities in different online platforms. Many of those have social network features, and thus are named social networking sites (SNS). SNS have become highly popular websites \cite{Mislove:2007}, with Facebook as the best exponent, with 1.4 billion active users. However, most SNS nowadays (e.g. Facebook, Twitter, Github) are data silos \cite{Yeung:2009}, where the user data is captured within ``walled gardens''. 	

A person may use a social code repository (e.g. Github\footnote{\url{http://github.com}}) to manage her own software projects, together with the ones from her organisation. In addition, she may be using a blogging platform to express her thoughts (e.g. Wordpress\footnote{\url{http://wordpress.com}}), a photo-sharing platform to publish her images (e.g. Flickr\footnote{\url{http://flickr.com}}), and an identity site (e.g. ORCID\footnote{\url{http://orcid.org/}}) to manage her online reputation. 

These platforms frequently allow other users to interact (like, cite, star) or republish the contents (blog posts, photographs, code, articles) to other social platforms. Besides, the blogging platform might be integrated with the photo service, in order to include images to ilustrate the articles, and last posts may appear in the digital profile. She might use the identy site to log into a manuscript submission and review site. This site might update the researcher's profile in ORCID when an article is accepted.

Interoperability describes the extent to which systems (in our case, SNS) can exchange data, and interpret that shared data, typically using some form of protocol or standard. Thus, we can find multiple levels of interoperability in the different SNS architectures. Some examples of these appear in user authentication, profile sharing/updating, resource distribution and activity distribution.

However, SNS interoperability could go far beyond that. In an ideal scenario, SNS would be able to communicate with one another seamlessly. For instance, we may think of the current academic ecosystem, where research institutions could re-structure their own social sites in order to enhance the collaboration among their members. 
In this scenario, they could be the primary SNS providers for their researchers, in a similar way they provide today with their primary email accounts. And the same as in e-mail\footnote{E-mail follows a federated architecture, i.e. it is decentralized, and no server have control over the whole ecosystem. Instead, any user can communicate with any other, regardless of where she has registered the e-mail account, as explained later.}, the institutions would like to be federated, in order to researchers be able to follow the work of their partners across their institutions. Moreover, researchers may want to move to a different institution, temporarily or permanently, and they would want to keep a compatible profile. Furthermore, there is an ecosystem of services that could benefit from this, including conferences, submissions sites or online learning platforms, who would obtain researcher's profile information easily.

In fact, the described scenario is the same path other internet technologies have already walked across. If we take a look at e-mail, for instance, we can see how proprietary ``walled'' solutions were the first ones to appear. These solutions did not interoperate among each other, i.e. a user could not send messages across different vendor solutions. Eventually, SMTP \cite{rfc812}, an open solution, emerged and dominated the market. Nowadays, users are able to send emails to other users, no matter which provider each user is actually using.

We refer to this as a ``federated'' solution, as providers are federated with each other (compliant with a common interoperable protocol specification) and the ecosystem is decentralized. Chao et al \cite{Chao:2012} describe
social networking federation as a paradigm where information on various social network systems can be seamlessly integrated in order to provide users a uniform and semantic view of their social connections. On the other hand, distributed services are those where data is scattered through the network nodes. Thus Federation is a subset of distributed, where nodes follow a server-client architecture, i.e. each server nodes manages the data of a set of client nodes. 
Datta et al \cite{Datta:2010} provide an extensive review of proposals of SNS on top of distributed systems such as peer-to-peer architectures and distributed hash tables.

Instant messaging is another example of successful federated technology. Every institution is now able to set up its own XMPP server \cite{rfc6120} and join this network of instant messaging (as Gmail/GoogleTalk did back in 2005). Thus, we believe the chance of ocurring the same with the social web is a plausible scenario. For instance, current successful vendor solutions nowadays (e.g. Twitter, Google+) do not allow establishing contacts among them. However, in the near future it may be possible to add contacts across networks and follow other users activity, no matter which the user affiliation is.

This vision of federated SNS, which obviously needs work on open federated protocols, have been extensively studied by the World Wide Web Consortium (W3C) during the last years. It started with the Social Web Incubator Group\footnote{\url{http://www.w3.org/2005/Incubator/socialweb/}}, whose main output was the complete report \textit{A Standards-based, Open and Privacy-aware Social Web} \cite{Appelquist:2010},  reviewed below. This group transitioned later to the Federated Social Web Incubator Group\footnote{\url{http://www.w3.org/2005/Incubator/federatedsocialweb/}}. They defined the social web acid test (SWAT) level 0 and 1 (SWAT0 and SWAT1), a set of use cases that social network platforms must fulfil in order to test their interoperatiblity. SWATs are not tied to any particular protocol or technology: they are top level actions. With the arrival of community groups to the W3C (more open than formal groups), the Federated Social Web Community Group\footnote{\url{http://www.w3.org/community/fedsocweb/}} was created. The topic was finally officially adopted by the W3C (i.e. not just as Incubator Group) and now it is being studied within the Social Web Working Group\footnote{\url{http://www.w3.org/Social/WG}}.

In this article, after reviewing the state of the art (section \ref{landscape}), we describe the paradigms that emerge in the interoperability among social networking sites (section \ref{paradigms}). Some of these paradigms are already described in Appelquist et al \cite{Appelquist:2010}. However, our analysis is made taking the social network framework described in Tapiador et al \cite{Tapiador:2012:SocialStream} as a basis. We believe such framework provides a solid base for understanding these interactions, with an approach which is clearer for (and closer to) software developers.

In section \ref{features}, a novel analytical framework for the interoperability of social networking sites is described. Three different kinds of actors are described, i.e. native, alien and foreign. The actor's SocialID is described, as a prerequisite for federated scenarios, and the different attributes of actors 
are analyzed from the federated point of view. Objects are described in the same way than actors, i.e. as native, alien and foreign. This section closes with a discussion about the federated behavior of several actions, i.e. authorship, ownership, reply, rating, mention and reshare.

In section \ref{analysis}, there is an analysis of seven representative technological solutions in the context of the federated social network framework: three federated frameworks present in literature, the user-centric identity framework OpenID, the APIs services that major SNS are providing to developers, the standard for federated social networks OStatus, and the federated software framework for real-time collaborative editing Apache Wave. This analysis shows how different technologies are solving the problem of interoperability and federated social networks.

Finally, section \ref{conclusions} gathers some concluding remarks and identified areas where there is room for improvement.

\section{The SNS Landscape}\label{landscape}
\subsection{Social Networking Sites}


Social network platforms have become among the most popular websites \cite{Mislove:2007}, although their history is rather short \cite{Boyd:2007}. There have been several attempts to define what constitutes a SNS.

Boyd and Ellison \cite{Boyd:2007} define them as web-based services that allow individuals to do three basic tasks, i.e. construct a profile, articulate a list of users with whom they share a connection, and view and traverse the list. Besides, they refer to other elements that SNS may have: avatars, privacy settings, customisation of relation names (beyond the "friend" clich\'{e}), posting resources to user’s ``wall'', private messages and any type of content sharing (photos, videos, etc). On the other hand, Kietzmann et al \cite{Kietzmann:2011} describe a framework with seven functional building blocks: identity, conversations, sharing, presence, relationships, reputation, and groups. Chinthakayala et al \cite{Chinthakayala:2014} perform a study focused in user experience, based on four criteria (navigation, interactivity, source credibility and intelligence). Tapiador et al \cite{Tapiador:2012:SNSFeatures} focus on the functional features of 16 platforms, highlighting the common ones (e.g. profile pages with avatars, or public sharing of comments) and analyzing the differences (e.g. privacy issues, or supported content types shared by users). Musia{\l} and Kazienko \cite{Musial:2013} provide in–depth analysis and classification of existing social networks (understood broadly, including email or instant messaging), according to a taxonomy of social networks.

Beyond mere feature description, Musia{\l} and Kazienko \cite{Musial:2013} describe a theoretical framework on the two main charateristics of SNS: the representation of social actors and their relations. According to the number of relations considered, they describe three kinds of social networks, i.e. homogeneous social network (HSN), system-based social network (SSN) and internet multisystem social network (ISN). They define internet identity as a digital, authenticable and permanent representation of a social entity, which may gather user identities from several sites. They also classify relations according to several criteria such as the consciousness of users or the visibility in the system, and describe the ``tie'' as a set of all relations that exist between two internet identities.

A social network framework is also introduced by Tapiador et al \cite{Tapiador:2012:SocialStream}. It presents a modular architecture for building any kind of web based social network site. The framework has two main components: objects and actors. Objects are understood as any kind of content managed in the SNS, such as documents, pictures, audio, video or events. On the other hand, actors are objects representing social entities (such as users, groups or organisations). Actors possess three additional characteristics. First, they have a \emph{profile}, with attributes such as location, webpage or an avatar that represents the actor. Secondly, they are able to build and maintain \emph{their social network}, establishing relations among them. Finally, they are able to \emph{perform actions on other objects}, including other actors. These actions include authoring, modifying, rating, following other objects. Actor's actions generate the timeline, which is a compilation of the actions performed by them. Access control is modeled through audiences, which set which actions (accessing, authoring, modifying, rating.. ) can be performed by other actors in the objects they own.

\subsection{Interoperability Issues}

There is significant work on the different issues around social network interoperability.

\subsubsection{Identity}

OpenID \cite{Recordon:2006a} emerged as a user-centric identity framework for the Web2.0. It specifies how web sites (named relay parties, RP) can authenticate users based on a URL, instead of the usual login and password credentials. It was developed as a solution to the problem of holding multiple login and passwords: with the advent of the Web 2.0 and the participation of the users in numerous sites, every user needs to create and maintain an account for each website where (s)he wants to participate in. OpenID proposes that RPs relay authentication in the identity provider (IP), a web service that authenticates the user, providing her with an identity URL, called OpenID identifier. This unique identifier will be used by the user for the rest of the web sites where she wants to log in. OpenID was pretty successful and it was adopted by multiple web sites. Several major providers became OpenID providers, including Google, Yahoo, Myspace, Wordpress and AOL. Plenty of users had OpenID identifiers, although many never noticed. OpenID says nothing about whether a web page must have any data attached to it or otherwise. In fact, those very concerned with anonymity may prefer to use opaque identities. However, there are other cases where people do maintain a primary source of authoritative information about their personas. E.g., a user may use her personal website as their profile page; perform comments and posts in other websites logged-in with her OpenID; and others may reach their profile through those comments \cite{Tapiador:2011:OpenIDSurvey}.

Although OpenID has interesting characteristics, such as universal identifiers and identity information discovery, it was quickly replaced by OAuth2, which has become the standard basis for authentication services \cite{Tapiador:2012:SNSAPIs}. OAuth is an authorization protocol for APIs. However, it is used as a single-sign-on solution. Users accept third-parties accessing their profile information in identity providers such as Google, Facebook or Twitter. This way, they are authenticating themselves, using the well-known "Login with Facebook" button.

WebID \footnote{\url{http://webid.info/}} is the wager from the Semantic Web, but it has not got traction so far. It is based on client-side certificates and the semantic vocabulary FOAF, described below.

Another solution designed for identity discovery is Webfinger \cite{rfc7033}. It is a protocol using email-like URIs, such as \textit{user@example.net}. It is inspired by the old UNIX finger service, which allowed to get information about a user in any computer that ran the service. Webfinger is based on the Web Host Metada specification \cite{rfc6415}, a method for retrieving well-known information on a web site, such as the site authors or copyright licenses. Webfinger resources are described as URI templates \cite{rfc6570}. 
Webfinger is used by the federated social networks GNU Social \footnote{\url{https://gnu.io}} and Diaspora \footnote{\url{https://joindiaspora.com}} to discover users on federated nodes.

Recently, a solution emerged aiming to compile the advantages of several of the former protocols: the third generation of OpenID technology, named OpenID Connect. It consists in an authentication layer on top of the OAuth2 authorization framework, which uses Webfinger as the discovery protocol. Although its specification was finished on Feb 2014 and its certification program launched on April 2015, it has already been adopted by major provides such as Google or Microsoft.  

\subsubsection{Profiles}

There is extensive work on identifing users whose profiles are scattered across SNS, matching the different profiles  \cite{Cortis:2013,Raad:2010,Jain:2013,Bennacer:2014,Chung:2014,Kong:2013}. Rowe \cite{Rowe:2009} shows how to build a distributed social graph by extracting identity information from 3 SNS (Facebook, MySpace and Twitter), converting and using OpenID to match the identities. Passant \cite{Passant:2008} shows how to export identity information from Flickr using their API and converting it to FOAF. Friend Of A Friend (FOAF) is an onthology describing persons and their relationships built on the top of RDF and OWL \cite{FOAF:2012}. People are able to export their social data in a distributed way, so a machine can collect the data and use FOAF profiles to find the list of all people two friends know, for example. Portable Contacts (PoCo) \footnote{\url{http://portablecontacts.net/}} describe a way to access user's address books and friend lists. They define a common access pattern and contact schema for retrieving both the social network and the contact details.

\subsubsection{User Activities}

Among the solutions related to user activites, there is PubSubHubBub (PuSH)\footnote{\url{https://code.google.com/p/pubsubhubbub/}}, a protocol for subscribing to web feeds. It aims to solve the inefficient pattern associated with web syndication. Web feed consumers that are subscribed to web feeds need to periodically request the feed to check if there are new updates. PuSH enables these consumers to subscribe to updates providing a public URL to the PuSH hub. When the website has a new update, it pings the hub server, which fetches the new feed and multicasts all the subscribers with the updates. Subscribers can discover the PuSH hub address as a link entry in the feed.

Also related with feeds, Salmon is a message exchange protocol running on the top of HTTP \footnote{\url{http://www.salmon-protocol.org/}}. Its goal is decentralizing comments and annotations notifications made against Atom feed articles, such as blog posts. It uses the Atom format along with its threading extensions to describe the notification content, its author and related information. Salmon notifications are base64 encoded and signed to assert the authenticity of the origin.

\subsubsection{SNS APIs}

Regarding SNS APIs, Ko et al \cite{Ko:2010} review social network connect services from Facebook, Google and Myspace. They show how third-party websites leverage these services to offer social enhanced value to their users, without the need to build their own social network. Connect services provide third-parties with easy sign-in and enrich user data and experience by mashing up their own data with the pieces retrieved from the API, e.g. finding friends in the platform. On the other hand, this is also interesting for the provider that now introduces new kinds of activity to its streams. A formal survey of this connect services is presented by Tapiador et al \cite{Tapiador:2012:SNSAPIs}, showing OAuth as the basis for authentication services, JSON as the data format in APIs, and a lack of semantic standard representation for common data such as user profiles. A failed attempt of general standard APIs formalization for social networks was organized around OpenSocial \cite{Hasel:2011}, without achieving an active reception from major SNS actors. In the last years, OpenSocial has been changing to embrace other technologies such as OAuth 2.0, Activity Streams and PoCo. It is based on well-known technologies such as HTML, XML, JSON and Javascript. A more successful reception was obtained by Activity Streams, \footnote{\url{http://activitystrea.ms/}}, an specification for the serialization of social activities that was adopted by IBM, Microsoft, MySpace and others, and it is now the basis of the W3C Social WG work.

\subsection{Federated Social Networks}

\label{federation}

There have been a wide range of proposals for a federated framework for the social web. D\"{o}rk et al \cite{Dork:2007} describe Atomique, a decentralized photo service using RSS feeds and Trackbacks. Its functionality is very basic and it is limited to content (images) aggregation. Chao et al \cite{Chao:2012} present a reference model and an application (named ``Linked.Twitter.In'') that federates two social networks. Gondor and Hebbo \cite{Gondor:2014} introduce SONIC. Its main contribution is the description of a DNS-like component for identity resolution and profile migration. However, it does not describe how actual communication between platforms is made. Tramp et al \cite{Tramp:2014} propose the architecture DSSN (Distributed Semantic Social Network) for an open, distributed social network, which is built solely on Semantic Web standards, i.e. WebID, FOAF, Semantic Pingback and PubSubHubbub. A similar architecture was proposed in OStatus \footnote{\url{https://www.w3.org/community/ostatus/}}, one of the most successful federation frameworks so far. Ostatus is an open standard enabling users in different social network platforms to follow each other. It uses several protocols for the different stages of federation, including Webfinger for discovering users from their IDs, Portable Contacts for describing user’s social network, PubSubHubBub (PuSH) for subscribing to users’ activity and Salmon for notification updates. However, OStatus was designed with public feeds in mind, and it is not suitable for settings where privacy restrictions are needed. OStatus was the basis of the most popular open alternative to Twitter, the microblogging platform identi.ca \footnote{\url{https://identi.ca/}}. And a slightly modified version was used by Diaspora \cite{Bleicher:2011}, the popular open alternative to Facebook. OStatus was recently replaced by Pump.io \footnote{\url{http://pump.io/}}, a federated framework in the spirit of OStatus, although with more modern protocols, such as OAuth and the recent JSON version of Activity Streams.

%

It is worth mentioning the family of solutions built on top of XMPP \cite{rfc6120}, the standard for instant messaging. XMPP already provides some interesting features, such as federated identity. Buddycloud \footnote{\url{http://buddycloud.com/}} is a federated social network built on top of XMPP. It introduces activity channels and a bridge from XMPP to the web. Apache Wave \cite{Weis:2011}, the distributed, near-real-time, rich collaboration platform uses XMPP to federate identities and content.

Most of these frameworks are gathered in \textit{A Standards-based, Open and Privacy-aware Social Web} \cite{Appelquist:2010}, a report by the W3C inital Social Web Incubator Group. Besides enumerating decentralized social networking frameworks, it addresses several topics, while describing a problem, a use case, and the relevant technologies offering a solution to that problem. Topics include identity, profile, social media, privacy, activity and emerging frameworks.







\section{Paradigms of Interoperability of Social Networking Sites}\label{paradigms}

Several patterns arise from the interoperability of SNS features. This section gathers patterns from the point of view of the components described in a social network framework \cite{Tapiador:2012:SocialStream}, i.e  at the level of actor identity, authentication and the construction of distributed profiles; at the level of the social network, how contacts and content can be exported to other networks; and the level of the activities, the actions that actors take on objects. Finally, the case of social network federation is analyzed.

\subsection{Actors: The Authentication \& Identity Fragmentation Problem}

One of the key functionalities for the interoperability of social network platforms is \textbf{identity} \cite{Musial:2013}. Each user is unique, with a set of characteristics that identifies her, together with a set of connections with other users. Identity has been intensively studied in social sciences by psychology, sociology, philosophy and anthropology \cite{Tajfel:2010}. In fact, identity is even becoming a keystone in the Web. Since the Web 2.0 realm is giving people the opportunity of taking several actions based on the individual (publication of news, photos, comments, personal information, etc...), identity becomes an essential issue, not only from a sociological, but from a technological point-of-view. Furthermore, trustability and credibility emerge as key issues as well.

Most SNS use as de-facto standard the ancient user and password combination for identifying users. An overwhelming amount of different services appeared after the Web 2.0 boom, and thus users have to manage not only a lot of different passwords for a variety of services, but a growing number of different profiles in series of platforms  around the world.

Recently, authentication solutions appeared, such as the already mentioned OpenID and OAuth. However, none of them has completely solved the full problem yet. In the case of OpenID, its popularity grew at the end of the last decade, but sites have gradually abandoned it \cite{gilbertson_openid_2015}. On the other hand, OAuth has become the current state-of-the-art solution for federated web authentication. However, it lacks of universal identifiers. It currently requires a sign-in button for initiating authentication, such as "\textit{Sign in with Facebook}". This requires the host to explicitly support providers. Users cannot authenticate using a different and independent identity provider that supports OAuth if the host has not registered it and made it available.

Besides the burden associated with credential management, the current situation involves a fragmentation of user activity along all the SNS. Because user \textbf{profiles} cannot be seamlessly integrated, there is not a coherence between different profiles created in every SNS. Furthermore, there is not a well-supported standarized way to combine digital contents created on every platform, unless web developers explicitely support one option. Still, the identity fragmentation scenario may also be desirable for privacy reasons\cite{roger_clarke_human_1994}. Sometimes we do not want our activities traced and bound across every place we log in. However, in other cases, specially when we want to build a coherent identity and reputation, such interoperability would make things easier. In these cases, from the user point-of-view, it would be desirable to have one single profile that could be validated against any service she would be accessing. This would be the case of the researcher profiles. There are some pieces of information, such as affiliation, position, contact email, or short bio, that we have to provide to every single new manuscript submission and review site we use. The profile could be obtained from an external source so it has not to be filled-in again and again. Furthermore, profile data could be synchronized when there are new attributes or changes in the status. Following the last example, researcher profiles could be automatically updated when an article is accepted. It could be linked to the profile in the research institution or to an independent researcher profile site.

\begin{figure}
\centering
\includegraphics[scale=0.5,trim={3cm 5cm 3cm 5cm},clip]{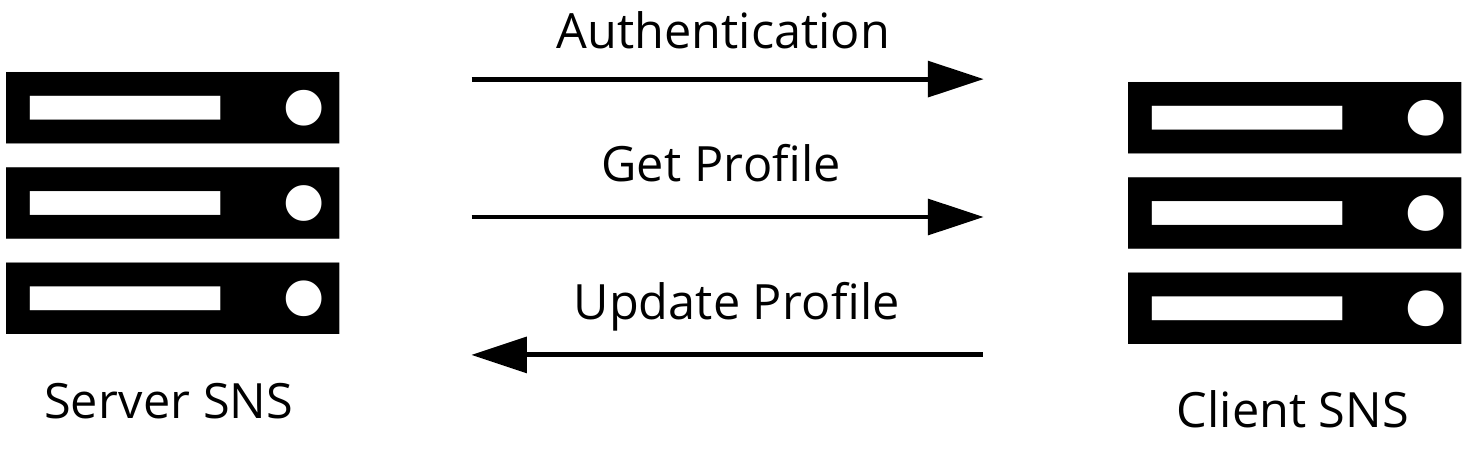}
\caption{Authentication and profile interoperability between SNS}
\label{fig:graphics:authorization-profile}
\end{figure}

Figure \ref{fig:graphics:authorization-profile} ilustrates a server SNS providing authentication services to a client site, as well as profile information about users. It also includes the synchronization of profile attributes, which happen in both ways: the server SNS notifies the client about new profile updates, and the client updates the server profile with a given interesting attribute, e.g current user's location.

\subsection{Contacts: Leveraging Social Connections}

Recent services provided by major social network platforms allow exporting the social graph. In other words, the \textbf{contacts} one actor has with other actors in the site can be retrieved by another web application.

This information is used by remote sites for customizing social user experiences based on people acquaintances \cite{Adams:2011}. We can find multiple examples, and some of them follow in order to illustrate the point.

We may think of the case of \textit{couchsurfing} services (e.g. couchsurfing.org, trustroots.org), a web page focused on offering users free hospitality services for their trips. Trust is an important issue when people are going to sleep with strangers. Thus, visitors will feel better if they find hints that the place they are visiting is safe. On the other hand, hosts may want information that indicates their guests are reliable. The couchsurfing service may grab contacts from both guests and hosts in the social network, so both sides can look at each other's connections. They may find some connections in common, which increases mutual trust (rather than hosting complete strangers).

Another example may be conference attendance. The conference scheduling service grabs the social graph from a social network site. This is a convenient way to know which conferences are attending those you follow. If you are following an interesting person in the social network, you may want to attend a conference this person is attending or speaking at.

\begin{figure}
\centering
\includegraphics[scale=0.5,trim={3cm 5cm 3cm 5cm},clip]{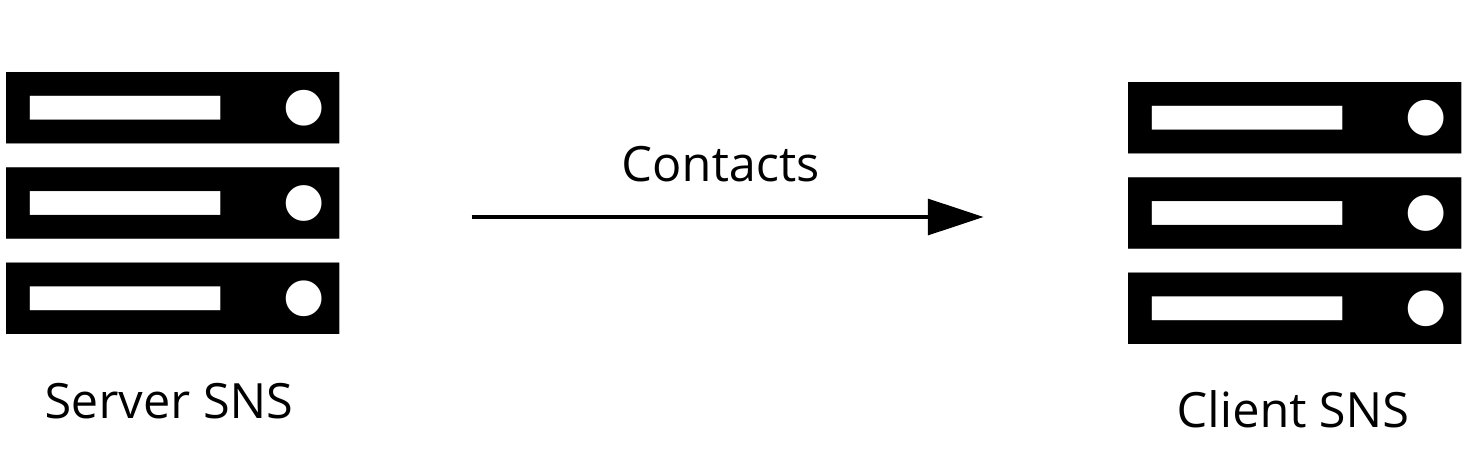}
\caption{Contact access between SNS}
\label{fig:graphics:contacts}
\end{figure}

Figure \ref{fig:graphics:contacts} shows the server SNS exporting the contact list or social graph of users to the client SNS.

\subsection{Content: Resource-Oriented Services}

There are other cases where the social sites may be interested in \textbf{content}, some specific resources created by the user. We introduce two examples of this pattern in the case of open source development and code repositories. 

Github \footnote{\url{https://github.com/}}, the popular social network for code sharing, can export the list of repositories that one person manages. Travis \footnote{\url{http://travis-ci.org/}} is a continous integration server \cite{Duvall:2007} that is in charge of checking the software application tests, and notify users when they have broken the application functionality in their last commit (i.e. source code change). Travis uses federated authentication to let users sign-in using their Github account. They grab the list of repositories one user has, allowing easy set up of repository testing. Finally, Github informs Travis when a repository changes, allowing Travis to update it and check the tests.

Another example is a photographers-oriented social network along with a printing service. A user may want to print some of the photos she manages in the social network. After being authenticated in the printing service, this service can grab the photographs associated to the user. Then, the user can easily select the appropriate ones to print.

\begin{figure}
\centering
\includegraphics[scale=0.5,trim={3cm 5cm 3cm 5cm},clip]{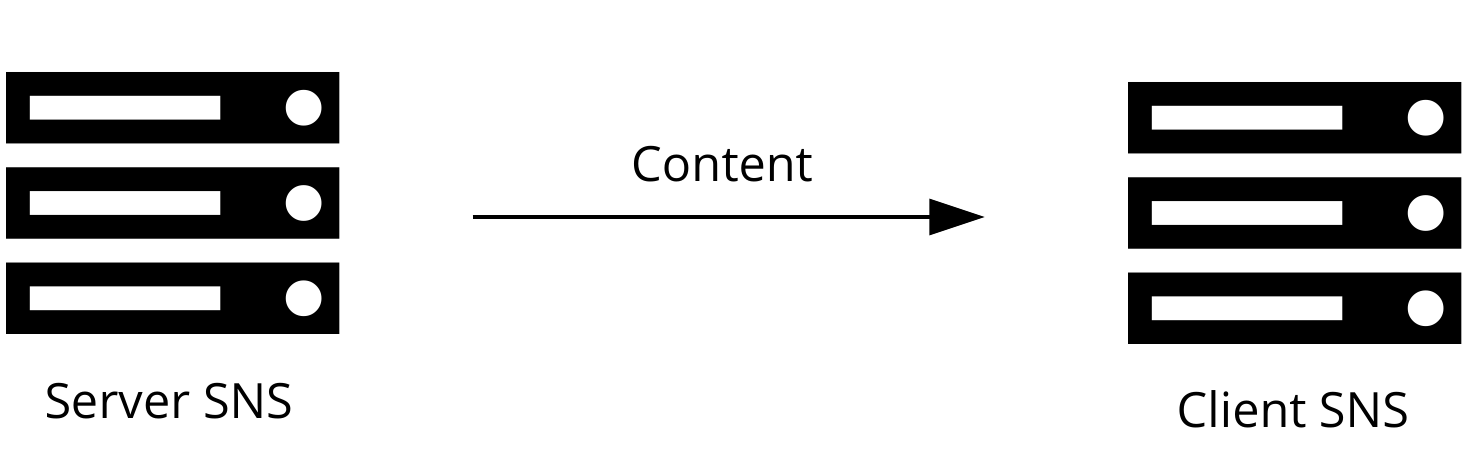}
\caption{Content access between SNS}
\label{fig:graphics:content}
\end{figure}

Figure \ref{fig:graphics:content} shows a SNS serving user's content collections to a client SNS. 

\subsection{Activities: SNS as the New Media}

Major SNS have become an outstanding communication media. Successful viral campaigns are pursued by marketers, and service providers encourage users to spread the word about them in their circles. In fact, it is expected that the stronger the ties, the higher the influence \cite{bond_61-million-person_2012}. 

Service providers implement buttons so it is easier for users to post to their social networks about a specific content/service, or report that they like it. Generic ``share'', \textit{Facebook's like} or \textit{Twitter's tweet} buttons have become very popular. They allow users to post specific content or perform a like action on a remote resource. Besides, SNS typically grab the user contacts in order to show them their friend liked content. This practice reinforces the positive feelings of users towards the product, because their friends like it too \cite{Adams:2011}

\begin{figure}
\centering
\includegraphics[scale=0.5,trim={3cm 5cm 3cm 5cm},clip]{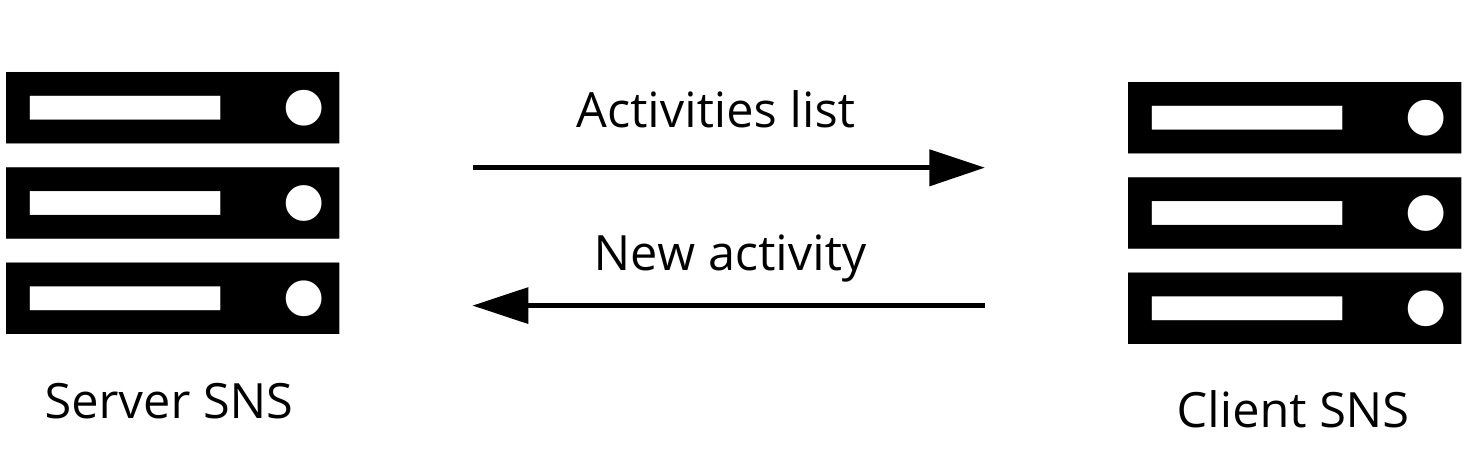}
\caption{Activity generation between SNS}
\label{fig:graphics:activity}
\end{figure}

Figure \ref{fig:graphics:activity} shows the client SNS grabbing user's \textbf{activities} from the server SNS. Besides, the client SNS creates new activities, such as posts or likes, in the server SNS.

\subsection{Federation: On Equal Terms}

Finally, there are cases where two institutions or networks interoperate on equal terms. This is the case among institutions such as universities or companies. They may want to share profile information, and their members may be interested in following each other. Contrary to former cases, there is reciprocity between both sites. The servers in this case must be interchangeable, so both of them have a role of client and server at the same time. This case reflects the case of \textbf{federation} among sites 

There are already protocols such as OStatus \cite{OStatus} that support this kind of interactions between SNS.

\begin{figure}
\centering
\includegraphics[scale=0.5,trim={2.7cm 5cm 2.7cm 5cm},clip]{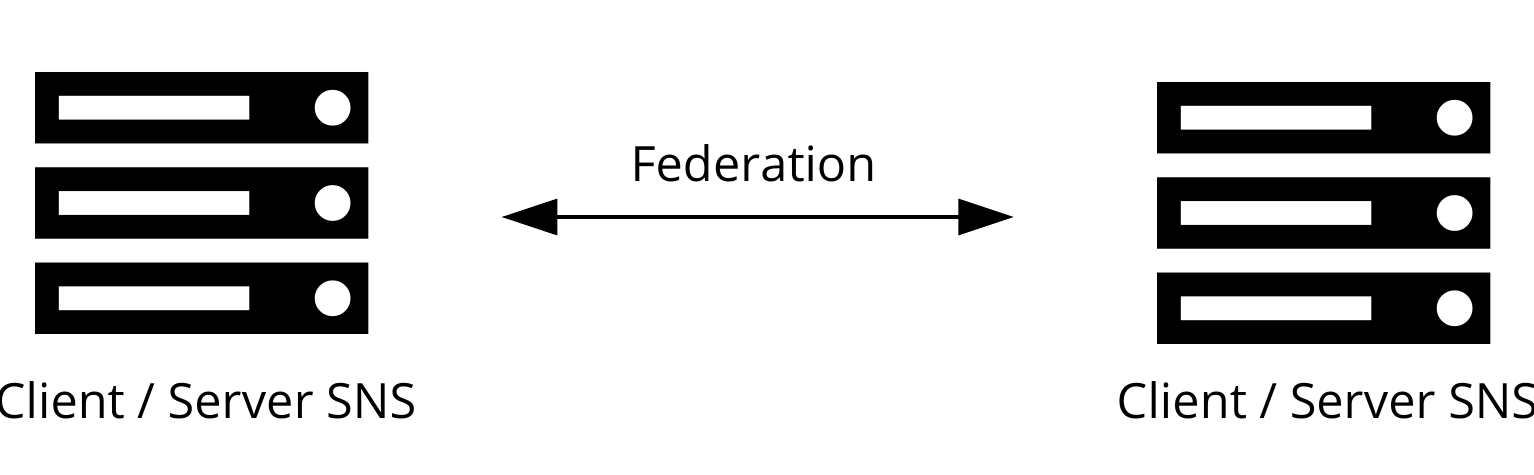}
\caption{Federation of SNS}
\label{fig:graphics:federation}
\end{figure}

Figure \ref{fig:graphics:federation} shows federation between SNS.

\section{Analytical Framework for the Interoperability of Social Networking Sites}

\label{features}

This section analyzes the cases described above and develops a analytical framework for federated SNS. This extends the frameworks described in \cite{Chao:2012}, \cite{Tapiador:2012:SocialStream} and \cite{Musial:2013} to include the interoperability features present in current SNS.

\subsection{Actors}

\textbf{Actors}, which represent social entities, are present in a federated fashion across SNS. Distributed identities agregated conform the virtual internet identity \cite{Musial:2013}. Actors were introduced to have three main characteristics, i.e. a \emph{profile}, with several attributes such as avatar, location, contact addresses; \emph{social relations} with other actors; and the ability to \emph{perform actions} on objects \cite{Tapiador:2012:SocialStream}. We will explore the federated requirements of these characteristics in each section below.

\subsubsection{Authentication}

Some kinds of actors, typically users, register and authenticate themselves into the site \cite{Joshi:2001}. Authentication is a main characteristic of their internet identity \cite{Musial:2013}. When users are authenticated, they can perform actions in the site on their behalf. Authentication can be federated beween sites using protocols such as OpenID or OAuth. According to their authentication mode, actors can be \textbf{native}, \textbf{alien} and \textbf{foreign}.

\begin{itemize}

\item \textbf{Native actors} are entities that are registered in a given SNS in the first place. An example of native actors are users that sign up into a SNS using one of the classic authentication methods, such as login and password. We call this server the \textbf{identity site}. The identity site can also be provider of an authentication service to other sites. In that case, users will be able to authenticate in remote sites, enabling the foreign actor case described below. Remote authentication is always performed after actors are authenticated in their identity site.

\item \textbf{Alien actors} are entities that are known to exist in an external identity site, but have never authenticated in a SNS. The other SNS, which has references to the alien actors, is called the \textbf{remote site}. An example of alien actors can be found in a site that analyzes the social network of a microblogging application, but does not support any kind of authentication. This is the case of many Twitter analysis tools that build statistics around Twitter accounts. Actors appearing in this site are alien actors. The actions performed by those remote actors in their identity site become remote actions in a remote site. Because these actions are not performed in the site, the remote actor's timeline must be filled with notifications from the identity site.

Figure \ref{fig:graphics:native-alien} shows a remote site with alien users that are referencing native users in their identity site.

Remote actors need an identifier for them to be discovered and referenced in remote sites. We call this identifier the social ID (see section \ref{sec:socialid}).

\begin{figure}
\centering
\includegraphics[scale=0.5,trim={3cm 5cm 3cm 5cm},clip]{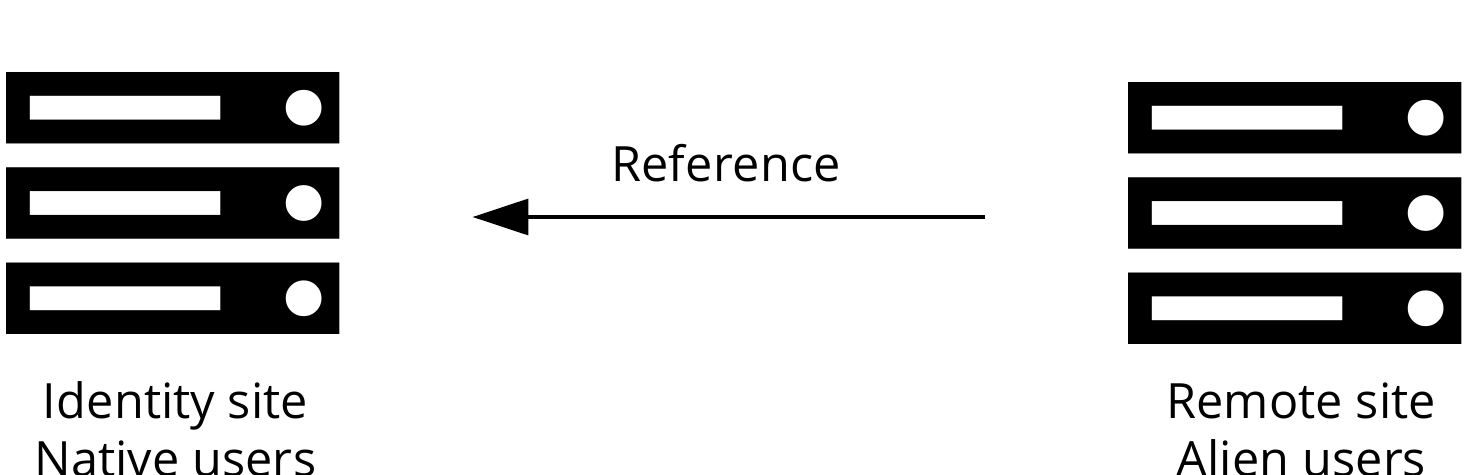}
\caption{Native and alien users in identity and remote sites, respectively}
\label{fig:graphics:native-alien}
\end{figure}

\item Finally, \textbf{foreign actors} are alien actors that use the federated authentication to login into the remote site. They gather most of the characteristics from both native and alien users. They are able to perform activities in their identity site as well as in the remote site they have signed in.

Remote sites may provide local authentication means as well, allowing their foreign actors to become native. An example of this proccess would be a remote site allowing their foreign actors to introduce and validate their email. Classical user and password authentication could be established afterwards using this mean. 

Figure \ref{fig:graphics:native-foreign} shows foreign users in a remote site that are authenticated by their identity sites, where they are native users.

\begin{figure}
\centering
\includegraphics[scale=0.5,trim={3cm 5cm 3cm 5cm},clip]{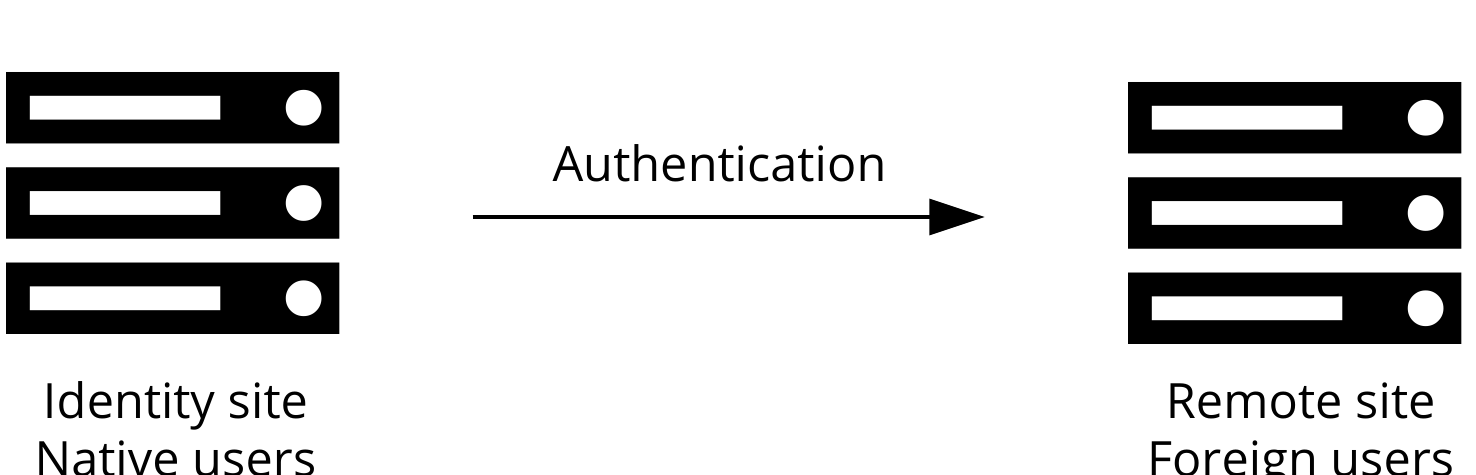}
\caption{Native and foreign users in identity and remote sites, respectively}
\label{fig:graphics:native-foreign}
\end{figure}

\end{itemize}

\subsubsection{Social ID \label{sec:socialid}}

A \textbf{Social ID} is mandatory in a distributed context. It is an identifier that is used to reference actors from any site in a federated social web. It represents the identity of a user across several systems.

Remote SNS are able to dereference and obtain information from the Social ID. This way, given a Social ID, users must be able to search and locate those actors from a given site in the federated system. Besides, Social IDs are useful for identifing actions performed by the same actor in different remote sites.

\subsubsection{Actors' Representation: Profile}

The profile is made up of data attached to an actor, such as a name, an avatar, contact emails, location address, description or personal websites. Remote sites query the identity site to obtain more information about an alien actor when they discover her Social ID. 

In addition, remote sites may subscribe to profile changes. When a profile is updated, the identity site notifies it to the remote sites that have been subscribed to it. This way, remote sites may have always fresh data about their alien actors.

The profiles may also be modified by remote sites. These might change some attributes directly. For example, a check-in application like Foursquare might change the current-location attribute of the users in a social network like Facebook. Moreover, some profile attributes might be updated indirectly, by the activities performed in remote sites, as explained in section \ref{sec:actact}.

\subsubsection{Actor Collections}

Actor's profiles may include collections attached to them. One of these are \textbf{content}, i.e. objects such as images, videos or code repositories. The typical collection gathers the content owned by the user. But there may be other collections, which are related to several actions, such as published articles authored by the user, favorite videos, etc.

Other collections include the set of actor's \textbf{contacts} o social ties \cite{Chao:2012} that define the actor's social network. They are available in interoperable SNS for retrievement. 
We are taking into account direct intentional relations here \cite{Musial:2013}, which result from the explict actions of actors aggregating other actors to their contact list. Regarding functional considerations of mutuality, relations can be unidirectional and bidirectional\cite{churchill:2005}. Unidirectional relations do not require a confirmation, but bidirectional ones do. In a distributed environment, unidirectional ``follow'' relations have a better fit, because they can be established between a native and an alien actor without any communication protocol among them. 

In the case of foreign actors, both identity and remote sites can communicate contact updates to each other, because the foreign actor is able to perform actions in the remote site, such as establishing new contacts. These contacts should be communicated to the user identity site, so the user contact list is updated.

\subsection{Objects}

Objects \cite{Tapiador:2012:SocialStream} or resources \cite{Chao:2012} are content managed by actors. Similarly to actors, objects may be native, alien or foreign. Native and foreign users are able to create \textbf{native objects} in a site, such as text, files, images, audios, videos or other kind of content. The server where the objects are created in the first place is called the \textbf{content site}.

Native objects may be referenced through notifications or activities from the content site to another site, named the \textbf{remote site}. For instance, we may consider the case where a native actor \textit{Bob} in site \textit{B} follows native actor \textit{Alice} in site \textit{A}. When \textit{Alice} posts a photo to site \textit{A}, this photo becomes a native object in site \textit{A}, which becomes the photo's content site. The notification about the photo creation should reach site \textit{A}'s alien actor \textit{Bob} in remote site \textit{B}. This notification should have a reference to \textit{Alice}'s photo in site \textit{A}. Therefore, each content object should have an \textbf{object ID} so it can be referenced from different SNS.

\textit{Alice}'s photo becomes an \textbf{alien object} in remote site \textit{B}, which will have information about the photo obtained from server \textit{A}. The remote site (site \textit{B}) may just reference the object or retrive a representation and cache it. The advantage of referencing is that there are not synchronization issues to be taken into account, less storage space is required, and user's browsers retrieve the object directly from the original site. This is, for instance, how Gravatar works (``Globally Recognized Avatar''). People upload their avatar to the Gravatar service, which is linked to an email. Other sites show their users' avatars just by referencing the URL in Gravatar's server. Browsers download the avatar directly from the Gravatar server. 

On the other hand, the advantages of caching is that it can be seamelessly integrated with the rest of the site (e.g. searched) and will be persistent if it is deleted in the content site. For instance, in the case of images, specific thumbnails can be generated, in order to improve efficiency. Objects may become obsolete or deleted in the content site, which might be able to notify the remote site on object changes. An example of the latter is how the Google Images search service work. It shows a cached thumbnail of each indexed image.

In the same way that actor's profiles, remote sites may subscribe to object updates, so the content site could notify remote sites on object changes. Then, the remote sites would be able to refresh the cached copy of the alien object.

\begin{figure}
\centering
\includegraphics[scale=0.5,trim={3cm 5cm 3cm 5cm},clip]{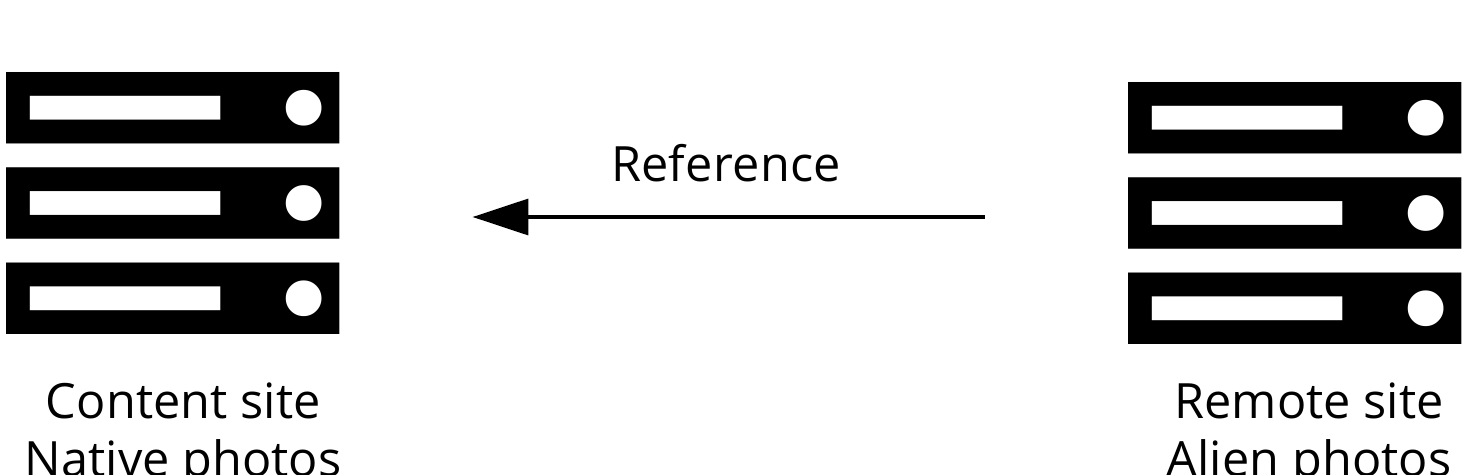}
\caption{Native and alien photographs in content and remote sites, respectively}
\label{fig:graphics:native-alien-content}
\end{figure}

Figure \ref{fig:graphics:native-alien-content} shows native photographs in their content site referenced as alien objects in a remote site.

Finally, a \textbf{foreign object} is defined as content that can be changed in the remote site. In the example, it could be possible that \textit{Bob} was able to modify the photograph, whose changes may reach back to site \textit{A}.

\subsection{Privacy Controls}

Users typically want to share different parts of their profile to different services. They may want to provide the minimum required information to some remote sites, but detailed information to other trusted remote sites.

Actors are also able to impose privacy controls to native objects. However, these objects may be cached in authorized remote sites. Remote servers should ensure that privacy restrictions are applied. In the example above, \textit{Alice}, a native user in site \textit{A}, may want to share her photo with \textit{Bob}, a native user in site \textit{B}, but not with \textit{Carol}, who is also a native in site \textit{B}. Site \textit{B} should ensure that \textit{Carol} is not able to retrieve \textit{Alice}'s photo. 

Ths is the same case with email. Email servers ensure that email messages sent to a user cannot be reached by others. In the literature, we can find several works addressing content re-sharing policies and their enforcement \cite{Squicciarini:2009,Appelquist:2010}

\subsection{Actions \& Activities}

Identity sites record actions from their native actors. These activities are usually available as feeds. Besides, a streaming API may allow remote sites to subscribe to activities \cite{Tapiador:2012:SNSAPIs}. 

In popular SNS, users may allow remote sites to post activities on their behalf. This is the case of foreign actors, where activities may be exchanged in both ways between identity and remote sites. On one hand, remote sites retrieve actor's activities. On the other hand, remote sites post new activities to the identity site. For example, when a Facebook user logs into a Couchsurfing server (i.e. a hospitality exchange web service), a foreign actor is created. The Couchsurfing site can grab some activities from Facebook to inform her hosts about the user's interests. If the user arranges a stay in a couchsufing house, the couchsurfing site could publish an activity in her Facebook wall as an announce of her future holidays. 

Activities are also a mean for content update. Identity severs may update the profile attributes or collections of their native users with the activities performed by actors in remote servers, such as contact establishment, posting, mentions, ratings, likes, etc. For example, \textit{Alice} starts following \textit{Bob}, who is a alien actor in \textit{Alice}'s server. This activity reaches \textit{Bob}'s identity server, who updates \textit{Bob}'s contact graph with \textit{Alice} as \textit{Bob}'s follower.

Objects can also be a source of activity streams. Different native and foreign actors may modify the objects, whose activities may be a vehicle for alien actors subscribed to object changes to discover and connect to them. Besides, there may be activity related to alien objects in the remote site, such as ratings, comments or mentions that could be propagated to the content site and update object's properties. For example, a photo in a content site could aggregate information about how many people liked the photo in other servers.

We extend here the examples of activities present in the Activity Streams specification \footnote{\url{http://activitystrea.ms/}} in order to include federated aspects:

\begin{itemize}

\item \textbf{Follow}: a native actor starts following an alien or foreign actor. The followed actor's identity site should be notified so the \textit{followers} collection is updated. This action may trigger a subscription from the remote site to the followed actor's activities published by her identity site.

The Follow activity have its opposite: Unfollow. The contact collection in the identity site is updated and subscription is cancelled.

\item \textbf{Authorship}: a native or foreign actor creates a new object in a content site. The new object can be accessed by her followers in remote sites. Alien actors may subscribe to the nofications of the new object.

\item \textbf{Ownership}: a new object is created in the wall of an actor. As in the case above, the activity may be propagated to the followers via the content and the identity sites.

\item \textbf{Reply}: an actor creates an object that is related with another one previously posted. Object authors may be notified and the comment may be agreggated to the content relation collection of replies in its content site.

\item \textbf{Rating}: the native actor rates an alien object from her identity site. As in the case above, object authors may be notified. The rating may be processed in the content site.

\item \textbf{Mention}: an alien actor is mentioned in the context of an object, e.g. tagged in a photograph. The actor may be notified in her identity site.

\item \textbf{Reshare}: the actor repeats a copy of an activity. Authors may be notified.

\end{itemize}


\section{Comparative Analysis of SNS Technologies}

\label{analysis}

The framework described above constitutes a powerful tool for analyzing interoperable and federation social network technologies. To ilustrate the point, the following technologies are examined in the context of the proposed framework (all mentioned previously in this paper): 

\begin{enumerate}

\item Atomique \cite{Dork:2007}, a decentralized photo service using RSS feeds and Trackbacks.

\item SONIC \cite{Gondor:2014} a solution for decentralized and heterogeneous federation of SNS.

\item DSSN, an architecture of a Distributed Semantic Social Network \cite{Tramp:2014}, built solely on Semantic Web standards.

\item The user-centric identity framework OpenID, together with the technologies that its use is associated with.
\cite{Tapiador:2011:OpenIDSurvey}

\item The services and formats that major SNS (e.g. Facebook, Twitter, Google+, Linkedin) are offering through their APIs \cite{Ko:2010,Tapiador:2012:SNSAPIs} 

\item OStatus \footnote{\url{https://www.w3.org/community/ostatus/}}, the protocol for social network federation

\item Apache Wave \cite{Weis:2011}, a federated software framework for real-time collaborative editing

\end{enumerate}

\subsection{Actors Support}

Actors are a basic feature in social networks. Native actors are present in all the technologies being analyzed. Alien actors are present as well, although the Social ID used to reference them varies. All the technologies analized already propose Social IDs (see section \ref{sec:socialid}). Atomique, OpenID and major SNS APIs use an HTTP URL (\textit{http://user.example.edu}). DSSN uses WebID's IRIs, the internationalized extension of URIs. 
Apache Wave uses the XMPP/email scheme (\textit{user@example.edu}). OStatus (through the Webfinger service) can use both. There is still an intense debate on the pros and cons of each scheme.

On the other hand, SONIC \cite{Gondor:2014} proposes the GlobalID, an alternative way to uniquely identify user accounts globally by using domain names. GlobalIDs are created using cryptographic hash functions and are dereferenced to LocalIDs using the Global Social Lookup System (GSLS) explained below.

Regarding foreign actors, only three technologies offer authentication services.  OpenID, as a user-centric identity framework, provides support for native and foreign actors. Native actors belong to the identity server that offers the OpenID service. Any other service supporting OpenID authentication has foreign actors. Users log in those services using the OpenID protocol, and are able to perform actions in the site.

On the other hand every SNS APIs in \cite{Tapiador:2012:SNSAPIs} are offering authentication services through OAuth APIs, and thus allowing remote sites to support foreign actors.

Finally, although it is not mentioned in the description of DSSN, WebID supports federated single sign-on on the basis of the WebID protocol.

\subsection{Actors' Representation: Profile}

\subsubsection{Profile Discovery}

Several technologies support discovering the profile from the Social ID. OpenID and OStatus support dereferencing the HTTP URL. OStatus supports the Webfinger protocol as well. DSSN supports this through WebID, while SONIC defines a complementary service: the Global Social Lookup System (GSLS), which is a directory service for the social records of all actors across the SNS. The GSLS provides the actual location of a social profile.

\subsubsection{Profile Representation}

Concerning the data format for profile information, identity sites export the actors' representation, in one or several formats: in the case of OpenID identifiers, including hCard and FOAF; DSSN uses WebID FOAF; major SNS APIs use JSON and XML; SONIC uses OpenSocial (which implies the use of JSON and XML); OStatus use Webfinger and HTML. Atomique and Apache Wave do not describe how actor profiles are represented.

\subsubsection{Profile Subscription} 

Only two of the analyzed technologies mention the posibility for other services to subscribe to profile changes. SONIC mentions that the GSLS supports this. DSSN proposes PuSH as a way to be subscribed to profile changes.

\subsubsection{Profile Updating} 

SONIC also mentions that profile attributes should be updated by other SNS, but does not mention how. DSSN proposes SPARQL update queries as a mean to update profile attributes. Finally, in popular SNS APIs, remote sites are usually able to set or update profile attributes from/to the identity site. 

\subsubsection{Profile Collections}

Regarding contact collections, every technology proposes a different format: SONIC relies on OpenSocial, OpenID profile pages use Microformats (XFN) and less often FOAF; DSSN uses just FOAF; OStatus uses Portable Contacts (PoCo); popular SNS APIs use XML and JSON. Atomique and Apache Wave do not specify how actor contacts should be exposed.

The analysis of OpenID identifiers show object collections described in different formats, such as HTML links and Microformats inside the profile. DSSN propose RDF as their representation format. OStatus uses Atom feeds and HTML for linking content collections inside the profile representation. Apache Wave retrieves the collection of waves each user is participating in. SONIC relies on OpenSocial. Atomique uses RSS feeds for photo collections. All these provide semantic description of the object collections, which facilitate the discovery of data (e.g. favorites, contacts) from the profile. This is not the case of major SNS APIs, which do not provide complete semantic descriptions and instead, the related collections are usually described offline, in the API documentation. This issue is well known, and breaks the REST principles \cite{Fielding:2000,Richardson:2007}. 

\subsection{Object Support}

Every technology supports native objects in their social networks. Object are also exported, which means that every technology supports alien objects. The Object ID is a URL, except in the semanic web framework DSSN, which uses IRIs.

\subsubsection{Object Representation}

Content sites export representations in several formats. Atomique uses RSS feeds; SONIC uses OpenSocial; DSSN uses RDF. In the case of OpenID identifiers, which are tied to the blogging community, blog posts are often exported as HTML pages and RSS / Atom feeds. Popular SNS APIs export content in JSON and XML. OStatus uses ActivityStreams, while Apache Wave uses the Wave XML format.

\subsubsection{Object Subscription \& Updating}

DSSN proposes PuSH as a way to be subscribed to profile changes, but does not describe how to update them.

Apache Wave and SNS APIs are the only technologies that describe support foreign objects. In Apache Wave, waves are created in their content server. When an alien actor is included in the wave, that wave is copied to the remote server, the alien actor's identity server. The alien actor is able to modify the wave, whose modifications come back to the original content server. This way, objects are modified both in the content and the remote server. The Wave protocol guaranties subscription to object changes and updates them automatically.

SNS APIs support object update, but do not describe a mechanism to subscribe to object changes.

The rest of technologies do not describe object subscription and update mechanism.

\subsection{Privacy Controls}

SNS APIs support OAuth scopes. This way, actors are able to impose privacy controls in their native objects. For instance, Facebook users, when using third-party apps, are typically asked to share part of their profile data (their friends, their wall notifications or their photos). Depending on the app, it requires different permissions, and the user might want to provide more data to some apps and less to others.

DSSN also proposes WebID access control, a mechanism for users to describe what data they want to share with other services.

The rest of technologies do not describe privacy controls.

\subsection{Actions \& Activities}

\label{sec:actact}

\subsubsection{Activity Feeds}

Almost every technology (excluding Apache Wave) lets services to fetch actor activities. Atomique uses RSS; SONIC relies on OpenSocial; DSSN and OStatus use ActivityStreams, OpenID identifiers use RSS and Atom, and popular SNS APIs use JSON and Atom.

Regarding object feeds, only DSSN and SNS APIs describe Atom feeds from the activities on object. Besides, the Wave Protocol includes a way to obtain activities on the waves.

\subsubsection{Activity Subscription}

DSSN, SNS APIs and OStatus are the only technologies that describe a method to subscribe to actor's activity updates: the PushSubHubBub (PuSH) protocol. Besides, SNS APIs provide streaming APIs, which may allow remote sites to subscribe to activities.

While DSSN and SNS APIs extend activity subscription to objects as well, OStatus only mention activities from actors. We should include the Wave Protocol as well, which supports subscription to activities on objects.

\subsubsection{Activity Notifications}

Every technology, except Apache Wave, describe a way to publish activity notifications to the actor's identity server. Atomique uses Trackback; SONIC relies on OpenSocial, DSSN introduces the Semantic Pingback, OpenID profiles are linked to plain Pingback; OStatus uses the Salomon protocol. Popular SNS describe several means of activity notifications. They even support users allowing remote sites to post activities on their behalf. This is the case in Facebook applications. Facebook goes one step further by leting remote sites to define new activity verbs in their application. Users must have granted permissions to this remote sites through a specific OAuth scope. This is the case of foreign actors, where activities may be exchanged in both ways between identity and remote sites. On one hand, remote sites retrieve actor's activities. On the other hand, remote sites post new activities to the identity site.

Regarding activity notification on objects, they are only described in DSSN, which uses Semantic Pingback; SNS APIs, which define their own API methods, and Apache Wave, which supports this functionality through the Wave Protocol.

\subsubsection{Native Content Changes}

All the technologies, except OpenID, describe how activity notifications change native content.

In the case of some federated SNS frameworks (i.e. Atomique, SONIC, DSSN) and OStatus identity servers may update the contact graph of their native users or attributes of native content with the activities performed by alien actors, such as contact establishment, mentions, etc. Atomique describes how group photo feeds are updated with user photos. SONIC and DSSN describe remote contact establishment and profile changes. OStatus, through Salmon notifications, lets the activity related to alien objects in the remote site, such as ratings, comments or mentions be propagated to the content site. Apache Wave changes native objects through the Wave Protocol.


\section{Discussion}

Table \ref{table:sns-technologies} shows how interoperability features are implemented in the 7 SNS technologies analyzed.

\begin{landscape}
\begin{table*}[ht]
\caption{
Comparison of different technologies according to their interoperable aspect, using the proposed framework for interoperability in SNS.\hspace{\textwidth}
\emph{Legend}: \ding{52}: native implementation, system dependant; (\ding{52}): taken into account in design, but further development needed; \ding{56}: not supported;\hspace{\textwidth} XXXXX: protocol used for supporting the feature.} 
\centering 
\begin{tabular}{|c|c|c|c|c|c|c|c|c|}

\hline 

\multicolumn{2}{|c|}{} & Atomique & SONIC & DSSN & \begin{tabular}[l]{@{}c@{}} OpenID \\ (+ associated) \end{tabular} & SNS APIs & OStatus & Apache Wave  \\ 

\hline 

\multirow{8}{*}{Actor} & Native & \ding{52} & \ding{52} & \ding{52} & \ding{52} & \ding{52} & \ding{52} & \ding{52} \\
\cline{2-9}
& \begin{tabular}[l]{@{}c@{}} Alien \\ (Social ID) \end{tabular} & HTTP URL & GlobalID & WebID IRI & \begin{tabular}[l]{@{}c@{}} OpenID \\ (HTTP URI) \end{tabular} & HTTP URI & \begin{tabular}[l]{@{}c@{}} Webfinger, \\ HTTP URIs \end{tabular} & \begin{tabular}[l]{@{}c@{}} XMPP \\ identifier \end{tabular} \\ 
\cline{2-9}
& \begin{tabular}[l]{@{}c@{}c@{}} Foreign \\ (Authentication \\ service) \end{tabular} & \ding{56} & \ding{56} & \begin{tabular}[l]{@{}c@{}} WebID \\ protocol \end{tabular} & OpenID & OAuth & \ding{56} & \ding{56} \\ 

\cline{2-9} 

& \begin{tabular}[l]{@{}c@{}c@{}} Profile \\ discovery \\ from Social ID \end{tabular} & \ding{56} & GSLS & WebID RDF & HTTP & \ding{56} & \begin{tabular}[l]{@{}c@{}} Webfinger, \\ HTTP \end{tabular} & \ding{56} \\
\cline{2-9}
& \begin{tabular}[l]{@{}c@{}} Profile \\ representation \end{tabular} & \ding{56} & OpenSocial & WebID RDF & \begin{tabular}[l]{@{}c@{}} hCard, \\ FOAF \end{tabular} & \begin{tabular}[l]{@{}c@{}} JSON, \\ XML \end{tabular} & \begin{tabular}[l]{@{}c@{}} Webfinger, \\ HTML \end{tabular} & \ding{56} \\ 
\cline{2-9}
& \begin{tabular}[l]{@{}c@{}} Profile \\ subscription \end{tabular} & \ding{56} & GSLS & PuSH & \ding{56} & \ding{56} & \ding{56} & \ding{56} \\
\cline{2-9}
& \begin{tabular}[l]{@{}c@{}} Profile \\ update \end{tabular} & \ding{56} & (\ding{52}) & SPARQL & \ding{56} & HTTP & \ding{56} & \ding{56} \\
\cline{2-9}
& \begin{tabular}[l]{@{}c@{}} Contact \\ collections \end{tabular} & \ding{56} & OpenSocial & FOAF & \begin{tabular}[l]{@{}c@{}} XFN, \\ FOAF \end{tabular} & \begin{tabular}[l]{@{}c@{}} JSON, \\ XML \end{tabular} & PoCo & \ding{56} \\ 
\cline{2-9} 
& \begin{tabular}[l]{@{}c@{}} Object \\ collections \end{tabular} & RSS feeds & OpenSocial & RDF & \begin{tabular}[l]{@{}c@{}} Feeds, \\ Microformats \end{tabular} & Custom & Atom feeds & Waves \\ 

\hline 

\multirow{4}{*}{Object} & Native & \ding{52} & \ding{52} & \ding{52} & \ding{52} & \ding{52} & \ding{52} & \ding{52} \\
\cline{2-9}
& \begin{tabular}[l]{@{}c@{}} Alien \\ (Object ID) \end{tabular} & URL & URL & IRI & URL & URL & URL & URL \\ 
\cline{2-9} 
& Representation & RSS & OpenSocial & RDF & HTML & \begin{tabular}[l]{@{}c@{}} JSON, \\ XML \end{tabular} & ActivityStreams & Wave (XML) \\ 
\cline{2-9}
& Subscription & \ding{56} & \ding{56} & PuSH & \ding{56} & \ding{56} & \ding{56} & \begin{tabular}[l]{@{}c@{}} Wave \\ protocol \end{tabular} \\
\cline{2-9}
& Update & \ding{56} & \ding{56} & \ding{56} & \ding{56} & HTTP & \ding{56} & \begin{tabular}[l]{@{}c@{}} Wave \\ protocol \end{tabular} \\

\hline 

\multicolumn{2}{|c|}{ Privacy controls } & \ding{56} & \ding{56} & \begin{tabular}[l]{@{}c@{}c@{}} WebID \\ access \\ delegation \end{tabular} & \ding{56} & OAuth & \ding{56} & \ding{56} \\

\hline 

\multirow{5}{*}{Activities}& \begin{tabular}[l]{@{}c@{}} Actor \\ feed \end{tabular} & RSS & OpenSocial & ActivityStreams & \begin{tabular}[l]{@{}c@{}} RSS \\ Atom \end{tabular} & \begin{tabular}[l]{@{}c@{}} JSON \\ Atom \end{tabular} & ActivityStreams & \ding{56} \\ 
\cline{2-9} 
& \begin{tabular}[l]{@{}c@{}} Actor \\ subscription \end{tabular} & \ding{56} & \ding{56} & PuSH & \ding{56} & \begin{tabular}[l]{@{}c@{}} PuSH, \\ Streaming \end{tabular} & PuSH & \ding{56} \\ 
\cline{2-9} 
& \begin{tabular}[l]{@{}c@{}} Actor \\ notification \end{tabular} & Trackback & OpenSocial & \begin{tabular}[l]{@{}c@{}} Semantic \\ Pingback \end{tabular} & Pingback & Custom & Salmon & \ding{56} \\ 
\cline{2-9} 
& \begin{tabular}[l]{@{}c@{}} Object \\ feed \end{tabular} & \ding{56} & \ding{56} & Atom & \ding{56} & \begin{tabular}[l]{@{}c@{}} JSON \\ Atom \end{tabular} & \ding{56} & \begin{tabular}[l]{@{}c@{}} Wave \\ protocol \end{tabular} \\ 
\cline{2-9} 
& \begin{tabular}[l]{@{}c@{}} Object \\ subscription \end{tabular} & \ding{56} & \ding{56} & PuSH & \ding{56} & \begin{tabular}[l]{@{}c@{}} PuSH, \\ Streaming \end{tabular} & \ding{56} & \begin{tabular}[l]{@{}c@{}} Wave \\ protocol \end{tabular} \\ 
\cline{2-9} 
& \begin{tabular}[l]{@{}c@{}c@{}} Object \\ notification \end{tabular} & \ding{56} & \ding{56} & \begin{tabular}[l]{@{}c@{}} Semantic \\ Pingback \end{tabular} & \ding{56} & Custom & \ding{56} & \begin{tabular}[l]{@{}c@{}} Wave \\ protocol \end{tabular} \\
\cline{2-9} 
& \begin{tabular}[l]{@{}c@{}c@{}} Native \\ content \\ changes \end{tabular} & \ding{52} & \ding{52} & \ding{52} & \ding{56} & \ding{52} & \ding{52} & \ding{52} \\
\hline
\end{tabular} 
\label{table:sns-technologies}
\end{table*}
\end{landscape}

It is relevant to highlight that the table shows an \textbf{overwhelming diversity of technologies for solving the same problems}. There is no consensus on how these should be solved, not even in the Social ID used. The only consensus that seems to arise is the use of IRIs as the identifier for object resources (as URLs are a subset of IRIs). Thus, we could say that \textbf{the solution for Object IDs are IRIs}.

All SNS technologies provide native support for native actors and objects, supporting the framework main components definition. All but one (OpenID) also support native content changes. These functionalities are internal and tied to custom implementations (databases, programming languages and business logic of SNS).

The Social ID (i.e. support of alien actors) is a feature that is also supported by all of them, although using a diverse set of technologies and standards. \textbf{HTTP URLs are the most common solution for Social IDs}.

\textbf{The representation of objects} managed in SNS, \textbf{and their collections} attached to user profiles, \textbf{are representative of current fragmentation} of technologies for interoperability among SNS. Although they are supported in every technology analyzed, none of them comes to an agreement with another to use the same solution.

A similar thing happens with profile representation and contacts collection. As they are common features needed for standard ``friending'', they are supported by most of them, except Atomique (basic SNS, focused on photo repositories) and Apache Wave (focused on collaboration, not social networking). And in fact, the spectrum of solutions for representing profile is astonishing: semantic languages such as Semantic Web's RDF and Microformats; markup languages such as XML and HTML; or data formats such as JSON. 

This fragmentation is also present in the features that are less supported by SNS technologies. 4 different technologies support profile discovery from a Social ID. However, each of them is proposing a different solution for the same task.

\textbf{There is little support for authentication services} (i.e. foreign actors). OpenID, OAuth and WebID are the only frameworks available. It seems like offering authentication services is not a popular feature in SNS frameworks. This is in contrast with the importance that authentication services are gathering in the world of SNS data silos. Major SNS are competing to become the next ID issuers, and each of them manages a different ecosystem of web sites that use their authentication services and orbitate around them.

Notification of activities generated by users, supported by all except Apache Wave, is another example of technology fragmentation. Every technology uses a different solution for this task, although Atom seems to be the base technology supporting it. PuSH is the technology for activity subscription, but it is only supported by 3 technologies. Object feeds, subscription and notification are more rare, just supported by 3 technologies.

\textbf{Privacy support for interoperable SNS is something yet to be explored in-depth}. Only two technologies provide solutions to this issue. Although privacy is an important concern in the world of SNS, interoperable technologies for social networks currently do not provide a strong solution for it. This might be related to the additional layer of complexity involved, as it is easier to solve the problem of interoperability when contents are public. It is an open question if privacy would come next.

Finally, \textbf{there is not much work in the field of data synchronization}. Subscription for profile changes, support for profiles updates, and in the case of SNS content, subscription for object changes and support for object updates. Just two technologies explicitly give support to each of these features in each row. This is quite intriguing, because direct and automatic synchronization of profile and object changes should be a way to maintain social network coherence through a federated system. This problem is partialy addressed by subscription to activities, which are a way to propagate profile and object changes. Synchronization among federated SNS has come through activities first.



\section{Concluding Remarks}

\label{conclusions}


This article describes 5 paradigms of interoperability across social networking sites. These correspond to the layers in which we can find interoperability, specially relevant when thinking about federation: actors, contacts, content, activities, and full federation itself.

Building on those, an analytical framework for interoperability is described. This framework is proven to be broad and powerful enough to analyze 7 representative technologies for interoperability, showing how different proposals provide solutions to different aspects within the paradigms described. 

The overall analysis exposes an overwhelming disparity and fragmentation in the solutions for tackling the same problems. This is shocking when considering the field is not that new, but it is still behaving as a young field where divergence and innovative solutions bloom, and thus remaining far from converging and consensus. Although there are a few solutions where this consensus is clear and are widely adopted (e.g. in object IDs), there are multiple basic issues that are far from being widely standardized (e.g. in profile representation).

Federation has reached success in different services (e.g. email, instant messaging) but it has not in SNS (yet?). It remains an open question if it is going to, or why it did not. Our analysis clarifies that the current technical approach has focused in using activities as the main vehicle to interoperate and synchronize SNS. However, there are other possible approaches yet to be explored, such as profile and content synchronization (i.e. like Apache Wave does with objects). There is also still room for improvement in the adoption of authentication services and, specially, in privacy controls. Although privacy is becoming a major issue of concern after Snowden revelations \cite{landau:2013}, the SNS are not yet adapting to this scenario. 

Although SNS will probably tackle these issues, after observing the ecosystem evolution, we can expect a wide diversity of solutions appearing in these under-explored areas. It is clear that SNS are still young, and are ready to innovate their way forward.


\bibliographystyle{spbasic}      

\bibliography{references}

\begin{thebibliography}{47}
\providecommand{\natexlab}[1]{#1}
\providecommand{\url}[1]{{#1}}
\providecommand{\urlprefix}{URL }
\expandafter\ifx\csname urlstyle\endcsname\relax
  \providecommand{\doi}[1]{DOI~\discretionary{}{}{}#1}\else
  \providecommand{\doi}{DOI~\discretionary{}{}{}\begingroup
  \urlstyle{rm}\Url}\fi
\providecommand{\eprint}[2][]{\url{#2}}

\bibitem[{Adams(2011)}]{Adams:2011}
Adams P (2011) Grouped: How small groups of friends are the key to influence on
  the social web. New Riders Press

\bibitem[{Appelquist et~al(2010)Appelquist, Brickley, Carvahlo, Iannella,
  Passant, Perey, and Story}]{Appelquist:2010}
Appelquist D, Brickley D, Carvahlo M, Iannella R, Passant A, Perey C, Story H
  (2010) A standards-based, open and privacy-aware social web. {W3C} incubator
  group report, W3C, http://www.w3.org/2005/Incubator/socialweb/XGR-socialweb/

\bibitem[{man Au~Yeung et~al(2009)man Au~Yeung, Liccardi, Lu, Seneviratne, and
  Berners-lee}]{Yeung:2009}
man Au~Yeung C, Liccardi I, Lu K, Seneviratne O, Berners-lee T (2009)
  Decentralization: The future of online social networking. In: In W3C Workshop
  on the Future of Social Networking Position Papers

\bibitem[{Bennacer et~al(2014)Bennacer, Nana~Jipmo, Penta, and
  Quercini}]{Bennacer:2014}
Bennacer N, Nana~Jipmo C, Penta A, Quercini G (2014) Matching user profiles
  across social networks. In: Jarke M, Mylopoulos J, Quix C, Rolland C,
  Manolopoulos Y, Mouratidis H, Horkoff J (eds) Advanced Information Systems
  Engineering, Lecture Notes in Computer Science, vol 8484, Springer
  International Publishing, pp 424--438, \doi{10.1007/978-3-319-07881-6_29},
  \urlprefix\url{http://dx.doi.org/10.1007/978-3-319-07881-6_29}

\bibitem[{Bleicher(2011)}]{Bleicher:2011}
Bleicher A (2011) The anti-facebook. Spectrum, IEEE 48(6):54--82,
  \doi{10.1109/MSPEC.2011.5779793}

\bibitem[{Bond et~al(2012)Bond, Fariss, Jones, Kramer, Marlow, Settle, and
  Fowler}]{bond_61-million-person_2012}
Bond RM, Fariss CJ, Jones JJ, Kramer ADI, Marlow C, Settle JE, Fowler JH (2012)
  A 61-million-person experiment in social influence and political
  mobilization. Nature 489(7415):295--298

\bibitem[{boyd and Ellison(2007)}]{Boyd:2007}
boyd dm, Ellison NB (2007) Social network sites: Definition, history, and
  scholarship. Journal of Computer-Mediated Communication 13(1):210--230,
  \doi{10.1111/j.1083-6101.2007.00393.x},
  \urlprefix\url{http://dx.doi.org/10.1111/j.1083-6101.2007.00393.x}

\bibitem[{Brickley and Miller(2012)}]{FOAF:2012}
Brickley D, Miller L (2012) Foaf vocabulary specification 0.98. Namespace
  document 9

\bibitem[{Chao et~al(2012)Chao, Guo, and Zhou}]{Chao:2012}
Chao W, Guo Y, Zhou B (2012) Social networking federation: A position paper.
  Computers \& Electrical Engineering 38(2):306 -- 329,
  \doi{http://dx.doi.org/10.1016/j.compeleceng.2011.11.028},
  \urlprefix\url{http://www.sciencedirect.com/science/article/pii/S0045790611002047}

\bibitem[{Chinthakayala et~al(2014)Chinthakayala, Zhao, Kong, and
  Zhang}]{Chinthakayala:2014}
Chinthakayala KC, Zhao C, Kong J, Zhang K (2014) A comparative study of three
  social networking websites. World Wide Web 17(6):1233--1259

\bibitem[{Chung et~al(2014)Chung, Lin, Lin, and Cheng}]{Chung:2014}
Chung CT, Lin CJ, Lin CH, Cheng PJ (2014) Person identification between
  different online social networks. In: Proceedings of the 2014 IEEE/WIC/ACM
  International Joint Conferences on Web Intelligence (WI) and Intelligent
  Agent Technologies (IAT) - Volume 01, IEEE Computer Society, Washington, DC,
  USA, WI-IAT '14, pp 94--101, \doi{10.1109/WI-IAT.2014.21},
  \urlprefix\url{http://dx.doi.org/10.1109/WI-IAT.2014.21}

\bibitem[{Churchill et~al(2005)Churchill, Halverson et~al}]{churchill:2005}
Churchill EF, Halverson C, et~al (2005) Guest editors' introduction: Social
  networks and social networking. Internet Computing, IEEE 9(5):14--19

\bibitem[{Cortis et~al(2013)Cortis, Scerri, and Rivera}]{Cortis:2013}
Cortis K, Scerri S, Rivera I (2013) Techniques for the identification of
  semantically-equivalent online identities. In: Lacroix Z, Ruckhaus E, Vidal
  ME (eds) Resource Discovery, Lecture Notes in Computer Science, vol 8194,
  Springer Berlin Heidelberg, pp 1--22, \doi{10.1007/978-3-642-45263-5_1},
  \urlprefix\url{http://dx.doi.org/10.1007/978-3-642-45263-5_1}

\bibitem[{Datta et~al(2010)Datta, Buchegger, Vu, Strufe, and
  Rzadca}]{Datta:2010}
Datta A, Buchegger S, Vu LH, Strufe T, Rzadca K (2010) Decentralized online
  social networks. In: Furht B (ed) Handbook of Social Network Technologies and
  Applications, Springer US, pp 349--378, \doi{10.1007/978-1-4419-7142-5_17},
  \urlprefix\url{http://dx.doi.org/10.1007/978-1-4419-7142-5_17}

\bibitem[{D\"{o}rk et~al(2007)D\"{o}rk, N\"{u}rnberger, and
  Mart\'{\i}n}]{Dork:2007}
D\"{o}rk M, N\"{u}rnberger A, Mart\'{\i}n JV (2007) Atomique: A photo
  repository for decentralized and distributed photo sharing on the web. In:
  Proceedings of the 15th International Conference on Multimedia, ACM, New
  York, NY, USA, MULTIMEDIA '07, pp 152--153, \doi{10.1145/1291233.1291265},
  \urlprefix\url{http://doi.acm.org/10.1145/1291233.1291265}

\bibitem[{E.~Prodromou and Copley(2010)}]{OStatus}
E~Prodromou JW B~Vibber, Copley Z (2010) Ostatus 1.0 draft 2.
  \urlprefix\url{http://ostatus.org/sites/default/files/ostatus-1.0-draft-2-specification.html},
  http://ostatus.org/sites/default/files/ostatus-1.0-draft-2-specification.html

\bibitem[{Fielding(2000)}]{Fielding:2000}
Fielding R (2000) Architectural styles and the design of network-based software
  architectures. Doctoral dissertation

\bibitem[{Gilbertson(2011)}]{gilbertson_openid_2015}
Gilbertson S (2011) {OpenID}: {The} {Web}'s {Most} {Successful} {Failure}.
  Wired: Webmonkey
  \urlprefix\url{http://www.webmonkey.com/2011/01/openid-the-webs-most-successful-failure/}

\bibitem[{Gondor and Hebbo(2014)}]{Gondor:2014}
Gondor S, Hebbo H (2014) Sonic: Towards seamless interaction in heterogeneous
  distributed osn ecosystems. In: Wireless and Mobile Computing, Networking and
  Communications (WiMob), 2014 IEEE 10th International Conference on, pp
  407--412, \doi{10.1109/WiMOB.2014.6962203}

\bibitem[{Gregorio et~al(2012)Gregorio, Fielding, Hadley, Nottingham, and
  Orchard}]{rfc6570}
Gregorio J, Fielding R, Hadley M, Nottingham M, Orchard D (2012) {URI
  Template}. RFC 6570 (Proposed Standard),
  \urlprefix\url{http://www.ietf.org/rfc/rfc6570.txt}

\bibitem[{Hammer-Lahav and Cook(2011)}]{rfc6415}
Hammer-Lahav E, Cook B (2011) {Web Host Metadata}. RFC 6415 (Proposed
  Standard), \urlprefix\url{http://www.ietf.org/rfc/rfc6415.txt}

\bibitem[{Harrenstien and White(1982)}]{rfc812}
Harrenstien K, White V (1982) {NICNAME/WHOIS}. RFC 812,
  \urlprefix\url{http://www.ietf.org/rfc/rfc812.txt}, obsoleted by RFCs 954,
  3912

\bibitem[{H\"{a}sel(2011)}]{Hasel:2011}
H\"{a}sel M (2011) Opensocial: an enabler for social applications on the web.
  Commun ACM 54:139--144, \doi{http://doi.acm.org/10.1145/1866739.1866765},
  \urlprefix\url{http://doi.acm.org/10.1145/1866739.1866765}

\bibitem[{Jain et~al(2013)Jain, Kumaraguru, and Joshi}]{Jain:2013}
Jain P, Kumaraguru P, Joshi A (2013) @i seek 'fb.me': Identifying users across
  multiple online social networks. In: Proceedings of the 22Nd International
  Conference on World Wide Web Companion, International World Wide Web
  Conferences Steering Committee, Republic and Canton of Geneva, Switzerland,
  WWW '13 Companion, pp 1259--1268,
  \urlprefix\url{http://dl.acm.org/citation.cfm?id=2487788.2488160}

\bibitem[{Jones et~al(2013)Jones, Salgueiro, Jones, and Smarr}]{rfc7033}
Jones P, Salgueiro G, Jones M, Smarr J (2013) {WebFinger}. RFC 7033 (Proposed
  Standard), \urlprefix\url{http://www.ietf.org/rfc/rfc7033.txt}

\bibitem[{Joshi et~al(2001)Joshi, Aref, Ghafoor, and Spafford}]{Joshi:2001}
Joshi JBD, Aref WG, Ghafoor A, Spafford EH (2001) Security models for web-based
  applications. Commun ACM 44:38--44,
  \doi{http://doi.acm.org/10.1145/359205.359224},
  \urlprefix\url{http://doi.acm.org/10.1145/359205.359224}

\bibitem[{Kietzmann et~al(2011)Kietzmann, Hermkens, McCarthy, and
  Silvestre}]{Kietzmann:2011}
Kietzmann JH, Hermkens K, McCarthy IP, Silvestre BS (2011) Social media? get
  serious! understanding the functional building blocks of social media.
  Business Horizons 54(3):241 -- 251,
  \doi{http://dx.doi.org/10.1016/j.bushor.2011.01.005},
  \urlprefix\url{http://www.sciencedirect.com/science/article/pii/S0007681311000061},
  \{SPECIAL\} ISSUE: \{SOCIAL\} \{MEDIA\}

\bibitem[{Ko et~al(2010)Ko, Cheek, Shehab, and Sandhu}]{Ko:2010}
Ko MN, Cheek G, Shehab M, Sandhu R (2010) Social-networks connect services.
  Computer 43(8):37 --43, \doi{10.1109/MC.2010.239}

\bibitem[{Kong et~al(2013)Kong, Zhang, and Yu}]{Kong:2013}
Kong X, Zhang J, Yu PS (2013) Inferring anchor links across multiple
  heterogeneous social networks. In: Proceedings of the 22Nd ACM International
  Conference on Conference on Information \&\#38; Knowledge Management, ACM,
  New York, NY, USA, CIKM '13, pp 179--188, \doi{10.1145/2505515.2505531},
  \urlprefix\url{http://doi.acm.org/10.1145/2505515.2505531}

\bibitem[{Mislove et~al(2007)Mislove, Marcon, Gummadi, Druschel, and
  Bhattacharjee}]{Mislove:2007}
Mislove A, Marcon M, Gummadi KP, Druschel P, Bhattacharjee B (2007) Measurement
  and analysis of online social networks. In: Proceedings of the 7th ACM
  SIGCOMM conference on Internet measurement, ACM, New York, NY, USA, IMC '07,
  pp 29--42, \doi{http://doi.acm.org/10.1145/1298306.1298311},
  \urlprefix\url{http://doi.acm.org/10.1145/1298306.1298311}

\bibitem[{Musia{\l} and Kazienko(2013)}]{Musial:2013}
Musia{\l} K, Kazienko P (2013) Social networks on the internet. World Wide Web
  16(1):31--72

\bibitem[{Passant(2008)}]{Passant:2008}
Passant A (2008) : me owl: sameas flickr: 33669349@ n00. In: LDOW

\bibitem[{Paul M.~Duvall(2007)}]{Duvall:2007}
Paul M~Duvall AG Steve~Matyas (2007) Continuous Integration: Improving Software
  Quality And Reducing Risk. Addison-Wesley

\bibitem[{Raad et~al(2010)Raad, Chbeir, and Dipanda}]{Raad:2010}
Raad E, Chbeir R, Dipanda A (2010) User profile matching in social networks.
  In: Network-Based Information Systems (NBiS), 2010 13th International
  Conference on, IEEE, pp 297--304

\bibitem[{Recordon and Reed(2006)}]{Recordon:2006a}
Recordon D, Reed D (2006) Openid 2.0: a platform for user-centric identity
  management. In: Proceedings of the second ACM workshop on Digital identity
  management, ACM, New York, NY, USA, DIM '06, pp 11--16,
  \doi{http://doi.acm.org/10.1145/1179529.1179532},
  \urlprefix\url{http://doi.acm.org/10.1145/1179529.1179532}

\bibitem[{Richardson and Ruby(2007)}]{Richardson:2007}
Richardson L, Ruby S (2007) Restful web services, 1st edn. O'Reilly

\bibitem[{{Roger Clarke}(1994)}]{roger_clarke_human_1994}
{Roger Clarke} (1994) Human {Identification} in {Information} {Systems}.
  Information Technology \& People 7(4):6--37

\bibitem[{Rowe et~al(2009)}]{Rowe:2009}
Rowe M, et~al (2009) Interlinking distributed social graphs. In: LDOW

\bibitem[{Saint-Andre(2011)}]{rfc6120}
Saint-Andre P (2011) {Extensible Messaging and Presence Protocol (XMPP): Core}.
  RFC 6120 (Proposed Standard),
  \urlprefix\url{http://www.ietf.org/rfc/rfc6120.txt}

\bibitem[{Squicciarini and Sundareswaran(2009)}]{Squicciarini:2009}
Squicciarini A, Sundareswaran S (2009) Web-traveler policies for images on
  social networks. World Wide Web 12:461--484,
  \urlprefix\url{http://dx.doi.org/10.1007/s11280-009-0070-8},
  10.1007/s11280-009-0070-8

\bibitem[{Tajfel(2010)}]{Tajfel:2010}
Tajfel H (2010) Social identity and intergroup relations, vol~7. Cambridge
  University Press

\bibitem[{Tapiador and Carrera(2012)}]{Tapiador:2012:SNSFeatures}
Tapiador A, Carrera D (2012) A survey on social network sites' functional
  features. In: Proceedings of IADIS WWW/Internet 2012

\bibitem[{Tapiador and Mendo(2011)}]{Tapiador:2011:OpenIDSurvey}
Tapiador A, Mendo A (2011) A survey on openid identifiers. In: Next Generation
  Web Services Practices (NWeSP), 2011 7th International Conference on, pp 357
  --362, \doi{10.1109/NWeSP.2011.6088205}

\bibitem[{Tapiador et~al(2012{\natexlab{a}})Tapiador, Carrera, and
  Salvach\'ua}]{Tapiador:2012:SocialStream}
Tapiador A, Carrera D, Salvach\'ua J (2012{\natexlab{a}}) Social stream, a
  social network framework. In: International Conference on Future Generation
  Communication Technology (FGCT 2012)

\bibitem[{Tapiador et~al(2012{\natexlab{b}})Tapiador, S\'anchez, and
  Salvach\'ua}]{Tapiador:2012:SNSAPIs}
Tapiador A, S\'anchez V, Salvach\'ua J (2012{\natexlab{b}}) An analysis of
  social network connect services. In: Proceedings of WEBIST 2012

\bibitem[{Tramp et~al(2014)Tramp, Frischmuth, Ermilov, Shekarpour, and
  Auer}]{Tramp:2014}
Tramp S, Frischmuth P, Ermilov T, Shekarpour S, Auer S (2014) An architecture
  of a distributed semantic social network. Semantic Web 5(1):77--95

\bibitem[{Weis and Wacker(2011)}]{Weis:2011}
Weis T, Wacker A (2011) Federating websites with the google wave protocol.
  Internet Computing, IEEE 15(3):51--58

\end{thebibliography}

\end{document}